# A Quadrature VCO with Phase Noise Optimization for Wide Tuning Triple Band Frequency Generation

Hechen Wang, Fa Foster Dai, Feng Zhao,

Dept. of Electrical and Computer Eng., Auburn University, Auburn, AL 36849

*Abstract*— **This paper presents an analytical model on quadrature VCO (QVCO) phase noise performance as well as a wide-tuning triple band QVCO RFIC design. The phase noise model demonstrates the advantages of applying phase shift to quadrature coupling for QVCO phase noise reduction and bi-model oscillation elimination. As an example, we present a wide-tuning and low noise QVCO RFIC that contains a bottom-series QVCO with bipolar transistors for oscillation and NMOS transistors for coupling. The low-band and middle-band signals are generated from the QVCO outputs and its 2nd harmonics, while the high-band signal is obtained by frequency mixing. The tuning-range enhancement technique enables the triple-band frequency generation in the range from 1.4GHz to 8.7GHz without penalizing oscillator's phase noise. The proposed wide-band frequency generation RFIC is implemented in a 0.18 µm SiGe BiCMOS technology with 0.6mm² area. The measured phase-noise is -125.2, -119.5 and -108.8dBc/Hz at 1 MHz offset for the three bands, respectively.**

*Index Terms*— **harmonics, ISF, phase noise, quadrature VCO (QVCO), tuning range, triple band,**

## I. Introduction

Multi-standard communication transceivers require quadrature signals with low phase noise, large tuning range as well as accurate and deterministic quadrature phase relationships. Moreover, multi-phase clock generations are needed for emerging applications such as wireless communications [1], [2], frequency synthesis [3], [4] and data converters [5]-[7].

Quadrature signals can be generated by means of: (i) frequency dividers, which requires the voltage-controlled oscillator (VCO) to oscillate at twice or four times of the LO frequency and thus consumes more power; (ii), poly-phase filters, which is narrow-band and lossy and thus requires power hungry gain stages to amplify the weakened signals; and (iii) quadrature VCOs (QVCO) [8], which can achieve low noise and accurate quadrature phase across a wide frequency tuning range. This paper focuses on phase noise optimization for various QVCO topologies.

Phase noise of oscillators is well modeled by Leeson's equation and impulse sensitivity function (ISF) [9]. When quadrature cores are coupled with each other, noise generated from another core will contribute to the overall noise. However, ISF analysis pointed out that the noise penalty is the worst or most sensitive when the injected signal is 90 degree out of phase, namely, the maxima of the injected quadrature signal aligns with the zero-crossing of the in-phase signal. Therefore, it is worth investigating the effect of phase shift and coupling strength for quadrature signal coupling. In this paper, an analytical model is developed to intuitively explain the effect of phase shift on noise reduction. The analysis was compared with circuit level simulations and good agreements were obtained for different QVCO topologies.

A variety of coupling schemes can be applied to QVCO designs. Super-harmonic signals of the quadrature oscillators can be coupled using passive resonators [10]. This technique also helps minimizing the current source noise if proper phase shifting is implemented. Quadrature oscillator cores can also be coupled using capacitors [11], in which phase shift can be easily introduced to reduce the phase noise and avoid bi-modal oscillation. Without active coupling device noise, capacitive coupled QVCOs can achieve better phase noise performance. However this kind of QVCOs is normally weakly coupled, namely, the current used for coupling is less than the current used for oscillation. Transformer based QVCOs can end up with strong coupling [12], yet they suffer from significant area penalty. Alternately, quadrature oscillator cores can also be coupled using active devices. By connecting the coupling transistors in parallel or serial to the VCO switching transistors, parallel-coupled QVCO (P-QVCO) and serial-coupled QVCO (S-QVCO) can be formed. For optimal coupling, the P-QVCO appears to have better quadrature amplitude and phase matching. However, the S-QVCO achieves better phase noise [13]. In most of the P-QVCO and S-QVCO designs, the same type of transistors are used for oscillation and coupling. Following a hybrid S-QVCO architecture introduced in [14], we present a hybrid S-QVCO design that utilizes different types of transistors for oscillation and coupling to achieve the optimized phase noise performance with a wide tuning range.

This paper is organized as follows: Section II studied QVCO phase noise with ISF analysis approach and introduced an analytical model on the phase noise performances under different phase shift and coupling strength for various QVCO topologies. The analytical model was then extended to model the phase noise of P-QVCO and S-QVCO designs. As an example of S-QVCO designs, we discussed the advantages of using both bipolar and CMOS transistors for oscillation and coupling. Section III presents a design of a triple band frequency generator RFIC that utilizes a BiCMOS S-QVCO to achieve wide tuning range without penalizing its phase noise performance. Finally, measured results are given in Section IV and the conclusions are drawn in Section V.

## II. Phase Noise of QVCOs

To model QVCOs, let's analyze a generic QVCO structure that uses either passive or active devices for coupling as shown in Fig. 1 below. If capacitors are used for coupling, the QVCO is said to be a capacitive coupling QVCO (CC-QVCO). If transistors are used for coupling, the structure could be either a P-QVCO, where coupling transistors are in parallel with the oscillation transistors, or an S-QVCO, where coupling transistors are in series with the oscillation transistors. Each of the two VCO cores for quadrature generation consists of two cross-coupled transistors to provide negative $g_m$ to overcome the LC tank loss. The resonant frequency of a QVCO will deviate from the resonant frequency of its single VCO core due to the quadrature coupling mechanism stated in [18].

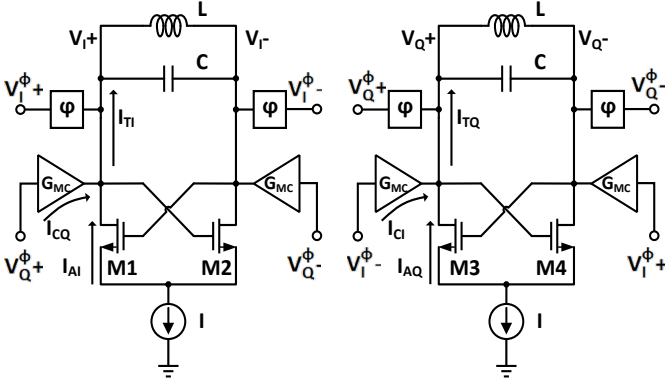

Fig. 1. A generic QVCO structure with coupling path delay φ and coupling factor *m*.

*A. QVCO ISF Simulation*

Impulse sensitivity function (ISF) analysis methodology is widely used in VCO designs to minimized oscillator's phase noise. Its concept can also be used to guide for low noise QVCO designs.

We simulated a QVCO and a single-phase VCO (SVCO) using a circuit similar to Fig. 1. The SVCO's results are obtained by setting coupling factor to 0 and observe one of the oscillation core, which shares the same simulation environment as the QVCO. By adding four identical ideal phase shifters into the coupling paths, the ISF of the SVCO and QVCO with 0º, 45º, 90º phase shift are also calculated using direct impulse response measurement method given in [9] and are shown in Fig. 2.

Since the power has been separated equally into two identical tanks, the ISF of QVCO without phase shift is halved comparing to the SVCO. With phase shift introduced into its coupling paths, the reduction grows till the phase shift reaches 90 degree. The maximum reduction ratio is 1/4 or -6 dB of the SVCO's ISF.

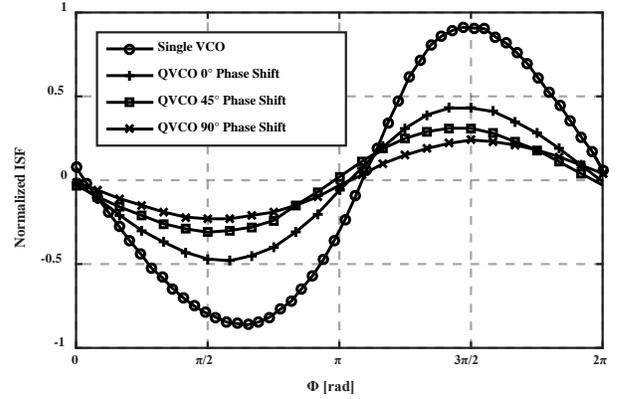

Fig. 2. Analytical and circuit simulation results on normalized QVCO phase noise PSD versus coupling factor *m* and phase shift.

According to this simulation, even without phase shift, QVCO's phase noise should be 3dB lower than that of the SVCO, which does not match with most of reported papers. The noise-modulating function (NMF), defined as the instantaneous drain current divided by the peak drain current over an output signal cycle, also contribute to VCO's phase noise. And the product of ISF and NMF, the effective ISF (ISF$_{eff}$), models VCO's phase noise more accurately.

Fig. 3 (a) shows the simulation results of SVCO's ISF, NMF and ISF$_{eff}$ together. The normal SVCO only have one current

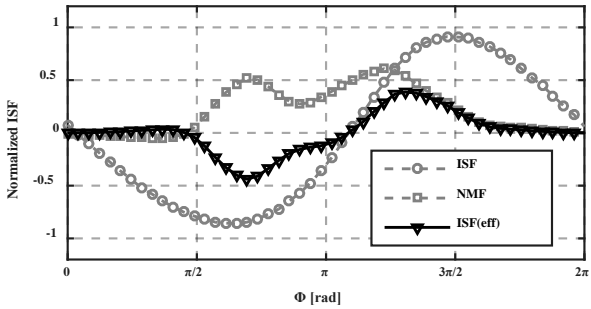

(a) ISF, NMF and ISF$_{eff}$ for SVCO.

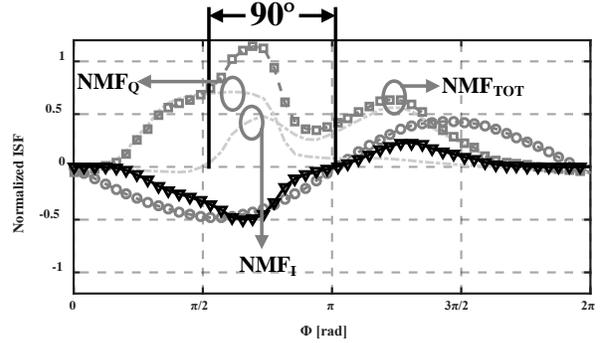

(b) ISF, NMF and ISF$_{eff}$ for QVCO with 0 phase shift.

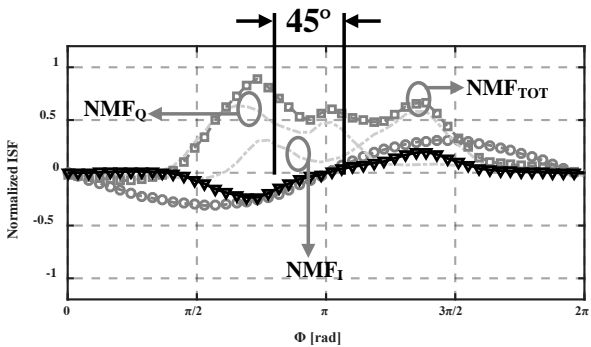

(c) ISF, NMF and ISF$_{eff}$ for QVCO with 45 phase shift.

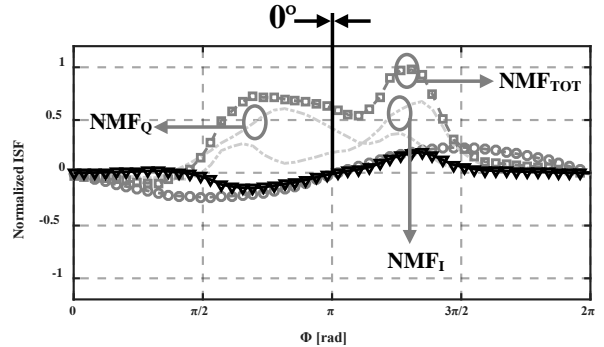

(d) ISF, NMF and ISF$_{eff}$ for QVCO with 90 phase shift.

Fig. 3. Simulated ISF, NMF and ISF$_{eff}$ comparison among SVCO and QVCO with different phase shift.

source and its NMF's central point is located around ISF's zero-crossing point. While the NMF of QVCO is modulated by both signals from I and Q cores. Fig. 3 (b), (c) and (d) shows QVCO's ISF, NMF and ISF$_{eff}$ with 0º, 45º, 90º phase shift respectively. There are two NMFs acting on one VCO core. NMF$_I$ is formed by its own current, NMF$_Q$ is formed by coupled core and NMF$_{TOT}$ is NMF$_I$ plus NMF$_Q$.

Without any phase shift in the coupling path, centers of NMF$_I$ and NMF$_Q$ are differed by 90º. The peak of coupled NMF$_Q$ is located at ISF's maximum point leading to a large ISF$_{eff}$. This explains the degradation of QVCOs without phase shift comparing with single phase VCOs. Along with the phase shift grows from 0 to 90º, the centers of NMF$_I$ and NMF$_Q$ overlapped with each other at ISF's zero-crossing point. Ideally, NMF$_{TOT}$ equals to single phase VCO's NMF, while the ISF of a QVCO with 90º phase shift is 6dB lower, which leads to a 6dB phase noise reduction between single phase VCO.

### B. Theratical Analysis on QVCO Phase Noise Model

In this section, we analyze the phase noise performance of different kinds of QVCOs. The phase noise power spectral density can be expressed using Lesson's equation as [17],

$$\mathcal{L}(\Delta\omega) = \frac{kTr_p}{V_0^2} F\left(\frac{\omega_0}{Q\Delta\omega}\right)^2 \quad (1)$$

where $K$ is Boltzman constant, $V_0$ is the oscillation voltage amplitude, $r_p$ is parallel resistance seen from the tank, $\omega_0$ is oscillation central frequency, $Q$ is loaded effective quality factor of the tank, $\Delta\omega$ is the offset frequency from the carrier frequency and the noise factor $F$ for a single-phase VCO is given by $F_{SVCO}=1+\gamma$. Obviously, part of the noise is contributed by the tank and another part comes from the oscillation transistors. Different QVCO structures will lead to different noise factor and output power, both of which will impact the overall phase noise.

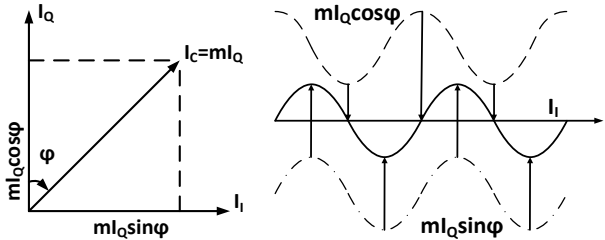

Fig. 4. Phasor diagram of current coupling in QVCO. Waveforms of the injected quadrature signal, decomposed into in-phase and quadrature-phase components for phase noise analysis.

Without losing the generality, let's analyze a generic QVCO structure as shown in Fig. 1. Assuming the coupling path introduces a phase delay of $\varphi$, the current coupled from the quadrature core $I_Q$ is rotated an angle $\varphi$ when it is injected into the in-phase core, as illustrated in Fig. 4, where $m$ is coupling factor defined as $m=|I_C|/|I_Q|$. The quadrature core current injected into the in-phase core is thus expressed as $mI_Q$. This current vector can be projected into the in-phase and quadrature phase axis with two perpendicular components: $mI_Q \sin\varphi$, which is in-phase with the in-phase core current $I_I$ and thus contributes to the signal strength, and $mI_Q \cos\varphi$, which is perpendicular to the current $I_I$ and thus contributes to the noise. This can be understood by observing that the in-phase component injects its maximum current at the peak of $I_I$ and thus has negligible noise contribution, while the quadrature phase component injects its maximum current at the zero-crossing of $I_I$ and thus contributes to the phase noise based on ISF analysis discussed in [9], shown in Fig. 4. Since $|I_Q|=|I_I|$, the total in-phase current becomes $I_C=(1+m\sin\varphi)I_I$. If the VCO output swing is not limited by excessively large $g_m$, the QVCO output swing will increase to $V_{QVCO}=V_0(1+m\sin\varphi)$. While the noise coupled from another core can be written as $\overline{V_{n-c}^2} = V_0^2(m\cos\varphi)^2$. The resultant noise factor of the QVCO is thus given by

$$F_{QVCO} = 1 + \left(\frac{m\cos\varphi}{1+m\sin\varphi}\right)^2, \quad (2)$$

where the first term is the noise factor of the single core and the 2$^{nd}$ term models the noise coupled from the quadrature core. With the modified noise factor of the QVCO, we can rewrite the phase noise power spectral density given in (1) as

$$\mathcal{L}(\Delta\omega) = \frac{kTr_p}{V_0^2(1+m\sin\varphi)^2}\left(1+\left(\frac{m\cos\varphi}{1+m\sin\varphi}\right)^2\right)\left(\frac{\omega_0}{Q\Delta\omega}\right)^2. \quad (3)$$

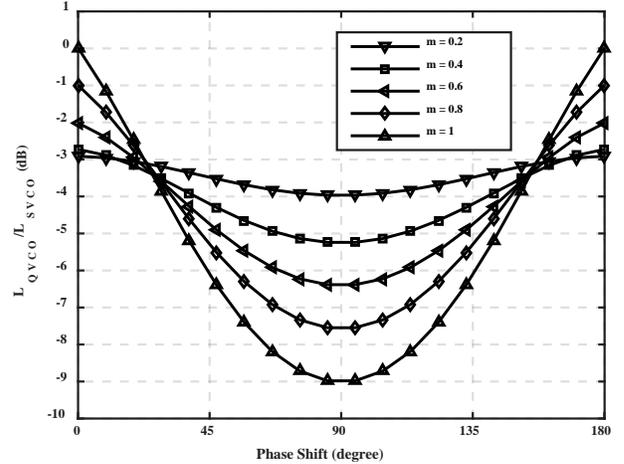

Fig. 5. QVCO phase noise PSD normalized to its SVCO counterpart versus coupling factor $m$ and phase shift.

Note that inserting a 90 degree phase-shift in coupling paths improves the phase noise by a factor of $(1+m)^2$. Fig. 5 illustrates the effect of QVCO phase noise reduction by introducing phase delays $\varphi$ under different coupling factor $m$. The QVCO phase noise is compared to its single phase VCO counterpart, assuming that the single-phase VCO counterpart flows the same bias current as that of each QVCO core, namely, the total QVCO power consumption is twice that of a single-phase VCO. For a strong coupling factor of $m=1$ and 90-degree phase shift in its coupling path, this leads to a phase noise reduction of 9dB when compared with a QVCO without phase shift. However, a QVCO with a coupling factor equals to 1 and a 90 degree phase shift in its coupling path can achieve 6dB better phase noise performance. However, large phase shift does not correspond to the best quadrature phase accuracy considering the mismatches [8]. To achieve good phase accuracy, phase shift of 45 degree and $m=0.5$ can be chosen which leads to phase noise reduction of 2.3dB. These observations were verified with measurements in [18].

The simulated phase noise results at 1MHz offset under different coupling factors and phase shift angles are shown in Fig. 6. The circuit level simulation results agree with the analytical results obtained using Eq. (3), which further confirm the correctness of the analytical model.

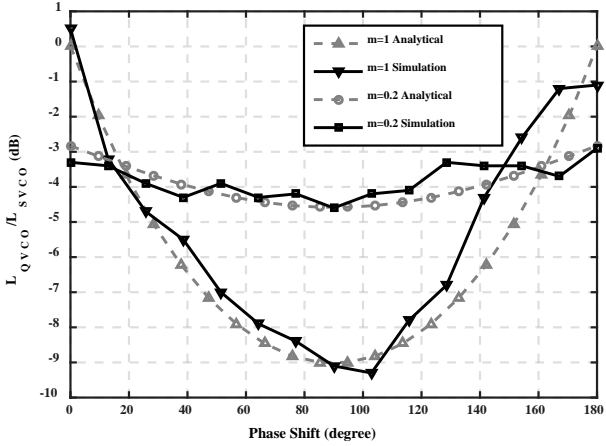

Fig. 6. Analytical and circuit simulation results on normalized QVCO phase noise PSD versus coupling factor *m* and phase shift.

In the above analysis, only tank loss is considered. To include noise from oscillation transistors, we can follow the approach used to obtain Eq. (3). Let's evaluate the noise factor for the in-phase core. The noise factor contributed by the tank and the oscillation transistors can be expressed as $1+\gamma$. This amount of noise is coupled from the quadrature phase core to the in-phase core, shown as a noise current at the in-phase core with the noise factor given as $(1+\gamma)(\frac{m\cos\varphi}{1+m\sin\varphi})^2$. Thus, the total noise factor including the noise contributions from both the tank and oscillation transistors can be modeled as

$$F_{QVCO} = (1+\gamma) + (1+\gamma)\left(\frac{m\cos\varphi}{1+m\sin\varphi}\right)^2, \quad (4)$$

where the first term $1+\gamma$ is the noise factor contributed by the in-phase core and the 2nd term represents the total noise coupled from the quadrature core. With some mathematical manipulations, the above term can be rewritten as

$$F_{QVCO} = \underbrace{1 + \left(\frac{m\cos\varphi}{1+m\sin\varphi}\right)^2}_{Tank\ Noise}$$
$$+ \underbrace{\frac{\gamma}{1+m\sin\varphi}\left(1 + \frac{m(m+\sin\varphi)}{1+m\sin\varphi}\right)}_{Oscillation\ Transistor\ Noise}. \quad (5)$$

With the modified noise factor, $F_{QVCO}$, phase noise power spectral density of the QVCO can be expressed as

$$\mathcal{L}(\Delta\omega) = \frac{kTr_p}{V_0^2(1+m\sin\varphi)^2}\left[\underbrace{1 + \left(\frac{m\cos\varphi}{1+m\sin\varphi}\right)^2}_{Tank\ Noise}\right. \quad (6)$$
$$\left.+ \underbrace{\frac{\gamma}{1+m\sin\varphi}\left(1 + \frac{m(m+\sin\varphi)}{1+m\sin\varphi}\right)}_{Oscillation\ Transistor\ Noise}\right]\left(\frac{\omega_0}{Q\Delta\omega}\right)^2.$$

## C. Phase Noise of P-QVCOs and S-QVCOs

Now, let's apply the above analytical results to two QVCO topologies: P-QVCO and S-QVCO, as shown in Fig. 7. Without additional phase shifters on the coupling paths, the P-QVCO has the following weakness: (i) only part of the bias current is used to create –gm for oscillation such that the signal power is reduced by a factor of $[1/(1+m)]^2$, and (ii) no phase shifting is introduced on the coupling path, namely, the quadrature-phase current is injected at the zero-cross of the in-phase current, leading to the maximal phase noise penalty. The thermal noises produced by both coupling and switching transistors are proportional to their gm. As a result, it is proportional to their (W/L) ratios and to their drain currents, assuming the same overdrive voltage is applied. Without the additional phase delay on the coupling paths, the phase noise of the P-QVCO can be modeled by simply inserting $\varphi=0$ into Eq. (6), namely,

$$\mathcal{L}(\Delta\omega) = \frac{kTr_p(1+m^2)}{V_0^2[1/(1+m)]^2}(1+\gamma)\left(\frac{\omega_0}{Q\Delta\omega}\right)^2 \quad (7)$$

where the term $[1/(1+m)]^2$ models the reduction of signal power due to slitting part of the bias current for coupling. As pointed out in [17], for a P-QVCO, phase noise and phase error are not independent of each other and both phase noise and phase error are strong functions of *m*. As the coupling between the cores is weakened, the phase error is worsened quickly. The phase accuracy requirement forces the P-QVCO design to choose the same size for both coupling and switching transistors. It is possible to improve the phase noise performance of the P-QVCO by weakening the coupling at the expense of its phase error performance [19].

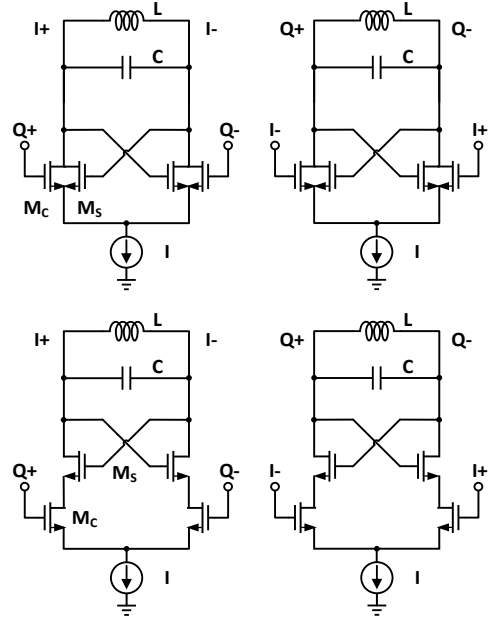

Fig. 7. Two commonly used QVCO topologies, (a) parallel coupled QVCO (P-QVCO) and (b) serial coupled QVCO (S-QVCO).

On the contrary, in the case of the S-QVCO, the phase error is almost independent of coupling factor *m*. In this case, the phase error is mainly dependent upon the amount of mismatches between ideally identical components. Thus, we can choose the transistor size independently to minimize the phase noise, while the phase error is not affected. Also, in an S-QVCO, the total bias current is used to generate the $-g_m$ for oscillation. Therefore, the signal power is not reduced unlike the P-QVCO case. Assuming that the size of the coupling transistors are chosen such that they contribute negligible noise to the tank, the phase noise of the S-QVCO is the same as a capacitively coupled QVCO except there is no phase delay on the coupling path. Inserting $\varphi=0$ into Eq. (6), we obtain the phase noise of the S-QVCO as

$$\mathcal{L}(\Delta\omega) = \frac{kTr_p(1+m^2)}{V_0^2}[1+\gamma]\left(\frac{\omega_0}{Q\Delta\omega}\right)^2. \quad (8)$$

To verify these analytical expressions, two circuit level models for a P-QVCO and an S-QVCO have been built and simulated in a CMOS process. The comparison between analytical and simulation results are given in Fig. 8. Phase noise performance comparison between P-QVCO and S-QVCO can be found in [13], where S-QVCO's phase noise is 146.1dBc/Hz at 3MHz offset, while P-QVCO's phase noise is 141.1dBc/Hz. The results agree with our analytical expressions.

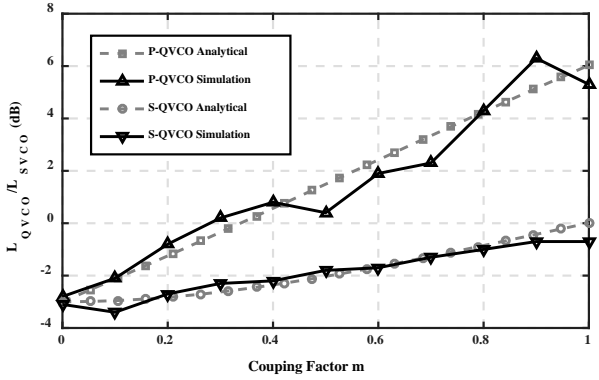

Fig. 8. Normalized phase noise PSD for P-QVCO and S-QVCO versus coupling factor *m*.

### D. Phase Noise and Tuning Range of S-QVCO Using BJT for Oscillation

As shown, S-QVCO has advantages over P-QVCO for phase noise reduction with flexibility to choose different types of transistors for coupling and oscillation. It's known that bipolar transistors (BJT) have better $g_m$ efficiency than MOSFETs. We thus discuss an S-QVCO structure that utilizes the BJTs for oscillation while the coupling is done through MOSFETs. As shown in Fig.9, NPN transistors ($Q_S$) form cross-coupled negative $G_M$ LC-VCOs, while coupling between the two VCOs is realized using four NMOS transistors ($M_C$). The oscillation NPN transistors can achieve high oscillation frequency and low phase noise with relatively low bias current, while the NMOS coupling transistors relax headroom requirement and provide better isolation.

Using BJT transistor for oscillation improves phase noise performance. The output equivalent noise current of a BJT transistor is given by

$$\overline{I_n^2} = \underbrace{2qI_C}_{Collector\ shot\ noise} + \underbrace{4kTr_bg_m^2}_{Base\ thermal\ noise}$$
$$= 4kTg_m\left(\frac{2qI_C}{4kTg_m} + r_bg_m\right). \quad (9)$$

For BJT transistors, we have $g_m=I_C/V_T$ and $V_T=kT/q$. Substituting them into above equation, we obtain $\overline{I_n^2}=4kTg_m(1/2+r_bg_m)$. In the proposed structure, NMOS transistors are used for coupling to reduce the required voltage headroom. The NMOS transistor's thermal noise, $\overline{I_n^2}=4kT\gamma g_m$, also contributes to the total noise, namely,

$$\overline{I_{n(BJT+MOS)}^2} = 4kTg_{m,BJT}\left(4kT\gamma\frac{g_{m,MOS}}{g_{m,BJT}} + \frac{1}{2} + r_bg_{m,BJT}\right). \quad (10)$$

Now we define the effective noise parameter $\gamma_{eff}=\gamma\left(g_{m,MOS}/g_{m,BJT}\right) + 1/2 + r_bg_{m,BJT}$. Hence, the noise factor of the hybrid S-QVCO can be expressed as

$$F_{S-QVCO}^{(BJT-MOS)} = 1 + \left(\frac{m\cos\varphi}{1+m\sin\varphi}\right)^2 \quad (11)$$
$$+ \gamma_{eff}\left[\frac{1}{1+m\sin\varphi}\left(1 + \frac{m(m+\cos\varphi)}{1+m\sin\varphi}\right)\right].$$

Comparing BJT and MOSFET's transcondctance $g_{m,BJT}=I_C/V_T$ and $g_{m,MOS}=\sqrt{2\mu C_{ox}(W/L)I_D}$, BJT produces higher $g_m$ than MOS transistors when the same amount of bias current is used. As a result, $g_{m,MOS} \ll g_{m,BJT}$, and the noise from the MOS transistor is negligible. The BJT oscillation pair works as a cascode buffer for the MOS pair, isolating their noise to the tank. Hence, $\gamma_{eff}$ becomes $\gamma_{eff} \approx 1/2 + r_bg_{m,BJT}$.

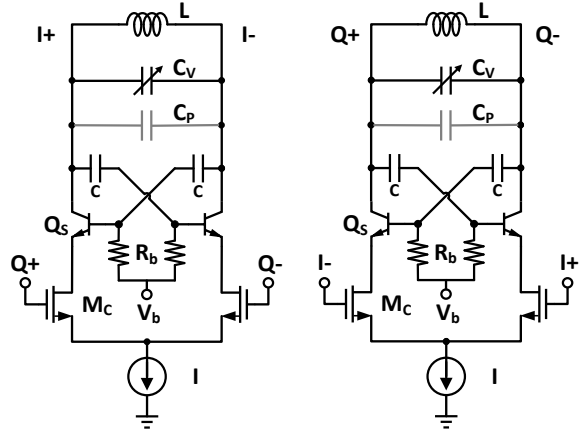

Fig. 9. Schematic diagram of S-QVCO using BJT for oscillation and NMOS for coupling.

To maintain steady-state oscillation, the cross-couple pair need to provide a transconductance $g_{m,BJT}=2/r_P$ to overcome the loss seen by the tank. Assuming $r_b=50\Omega$, comparing with tank resistance $r_p=L\omega_0Q=1nH\times 5GHz \times 2\pi\times 20=200\pi\Omega$, we notice that the best $\gamma_{eff}$ value is slightly larger than 0.5, which means 3dB lower phase noise than what an S-QVCO with MOS transistors for oscillation can achieve.

Using BJT transistor for oscillation also improves frequency tuning range (*FTR*). The oscillation frequency of the S-QVCO is given by $f_{osc}=(2\pi\sqrt{LC})^{-1}$, where *C* is the total capacitance seen across the inductors which includes the varactor capacitance $C_v$ and the fixed parasitic capacitance $C_p$. The VCO frequency tuning range (*FTR*) can be found as $FTR=C_p+C_{v,max}/C_p+C_{v,min}$.

The fixed parasitic capacitance $C_p$ consists of parasitic capacitance contributed by inductor, varactor, and all the transistors connected to the resonant tank. For high frequency oscillation, this fixed capacitance contributes to the overall LC-tank capacitance and thus limits the tuning range. The total fixed capacitance seen across the LC-tank for hybrid S-QVCO shown

in Fig 9 is given by [14]:

$$C_p = 2C_\mu + \frac{C_{gs}}{2} + \frac{C_\pi C_{gd} C_{gs}}{2(C_\pi C_{gd} + C_\pi C_{gs} + C_{gd}C_{gs})}, \quad (12)$$

As shown in the above equations, the use of MOS transistors for coupling in serial with BJT oscillation transistor pair reduces the fixed parasitic capacitance $C_p$ seen across the tank due to the increased capacitive degeneration ($C_{gs}$ and $C_{gd}$). This provides more tuning range and higher attainable oscillation frequency.

III. PROPOSED TRIPPLE-BAND FREQUENCY GENERATOR

Using a QVCO, we implement a triple-band frequency generator RFIC depicted in Fig. 10. The low-band quadrature signals are generated at the output of QVCO. The differential middle-band signals are obtained at the common-mode nodes of the cross-coupled differential pairs. Mixing the differential fundamental frequency signals ($f_0+$ and $f_0-$) with its second harmonics ($2f_0$) produces both upper and lower sideband ($2f_0 \pm f_0$) at output nodes of mixers. By filtering the lower sideband with a bandpass filter (BPF), signals with triple frequency $3f_0$ can be generated at the output of BPF. The BPF is formed by an inductor and two capacitors that resonants at $3f_{vco}$. As a result, triple-band signal generation is achieved with one QVCO and a mixer. The QVCO cores not only produce low-band frequency, but also works as a frequency doubler without extra circuit.

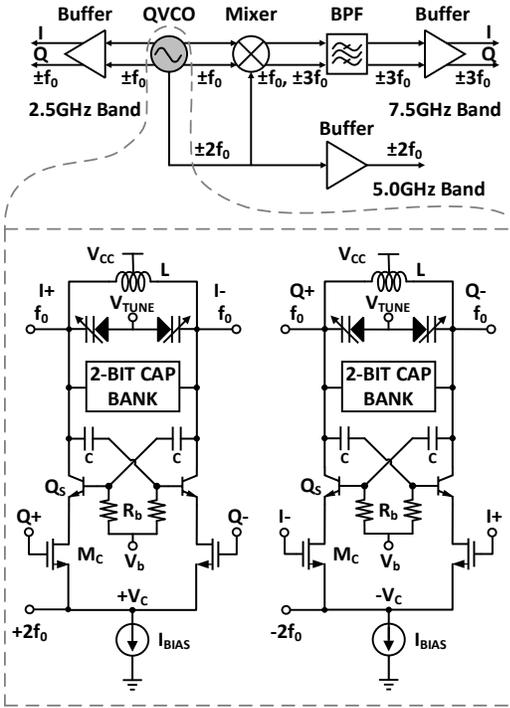

Fig. 10. Block diagram of the implemented triple band frequency generator and circuit diagram of the proposed hybrid BS-QVCO.

A. *Bottom-Series QVCO Core*

The implemented QVCO core circuit is illustrated in Fig. 10. SiGe bipolar transistors [20] are used for oscillation in order to obtain a better $g_m$ efficiency. Signal swing of the oscillator output was optimized to improve the phase noise performance. Meanwhile, if the oscillation BJT transistors are operated in saturation region, noise will be coupled through the parasitic PNP transistors to the substrate, causing degraded phase noise performance. A trade-off was made between maximizing the signal swing and avoiding BJT's saturation. This process involves careful design of balancing transistor size, bias voltage and bias current. The resonant tank is formed by a differential inductor with its peak Q factor optimized at 2.5GHz and two varactors with reasonable size to achieve targeted tuning range and required phase noise performance. This work uses bottom-series MOSFET coupled QVCO (BS-QVCO) to achieve low phase noise and lower power consumption.

B. *2-Bits Capcitor Array Band Switch*

As known, the tradeoff between tuning range and phase noise eventually determines the size of the varactors. Enlarge a varactor can increase VCO's tuning range, while its parasitic resistance will also increase and degrade Q factor of the tank leading to a worse phase noise performance. In order to widen the tuning range without affecting overall phase noise performance, this work employed a capacitor array switched by 2-bit control signals to switch the resonant frequency band instead of using large varactors. Binary weighted capacitance is adopted to enlarge the tuning range. Each switching bit controls two identical capacitors to compensate the process, voltage and temperature (PVT) variations. In this arrangement, two binary bits can provide the control of four different frequency sub-bands.

C. *Frequency Doubling & Tripling Schemes*

The double frequency band is obtained at the virtual ground nodes ($V_{C+}$, $V_{C-}$ in Fig. 10) of the differential pairs, where odd order harmonics are cancelled, leaving even order harmonics. Comparing to MOSFET transistors, which have larger parasitic capacitance, BJT differential pair has relatively larger 2nd order harmonic component at its common mode. Derived from differential pair nonlinear signal model, the 2nd order harmonic components shown at the virtual ground nodes of BJT and MOSFET differential pairs are given by

$$A_{V2^{nd},MOS} = \frac{I_{tail}}{V_{OV}^2/2} R_s, \quad (13)$$

$$A_{V2^{nd},BJT} = \frac{I_{tail}}{V_T^2} R_s, \quad (14)$$

where $I_{tail}$ is tail current of the differential pair, $V_{OV}$ and $V_T$ are transistor's overdrive voltage and thermal voltage, respectively; $R_S$ is the tail current source's output resistance. Thermal voltage is 26mV and MOSFET's overdrive voltage is around 100mV to 200mV for low power design, normally. Under the same tail current, $1/V_T^2$ is much larger than $1/(V_{OV}^2/2)$, which leads to a greater gain of the 2nd harmonic for BJT differential pairs. The amplitude of the 2nd order harmonics at the common mode node of a BJT differential pair can reach as high as 254mV$_{pp}$ or –5.95dBm under the operation condition.

As far as the triple frequency band concerned, a single balanced mixer [21] may cause large clock feed-through and coupling to the substrate. In order to achieve better spectral purity, our frequency tripler adopts the double-balanced topology, which has high common-mode rejection ratio and suppressed clock feed through. The triple frequency band generation can be formulated as: *Signal I*=$cos(\omega t)cos(2\omega t)=cos(3\omega t)$, *Signal Q* = $cos(\omega t + \pi/2)cos(2\omega t) = cos(3\omega t + \pi/2)$. Thus, mixing fundamental quadrature frequency signals with double-frequency signals generates quadrature phase triple-frequency tones.

## IV. IMPLEMENTATION AND MEASURED RESULTS

The triple-band frequency generator prototype is implemented in a 0.18μm SiGe BiCMOS process [22]. The QVCO cores consume 12mW and the mixer consumes 31mW (exclude output buffer power) under 2.5V supply voltage. Die photograph is shown in Fig. 11. QVCO with pads and output buffer occupies 1.8x1.0 mm$^2$, and the rest of the area is taken by the built-in band-pass filter for 7.5GHz output. The QVCO core area is 1.2x0.5 mm$^2$.

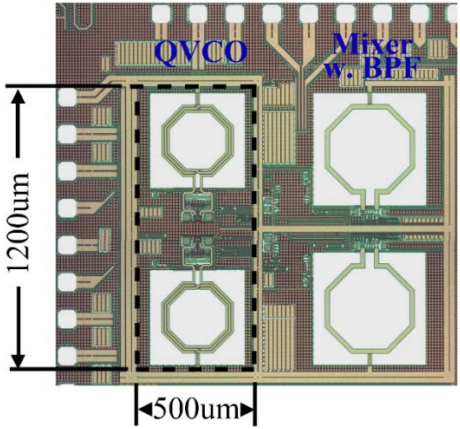

Fig. 11. Die photo of the triple band frequency generator RFIC.

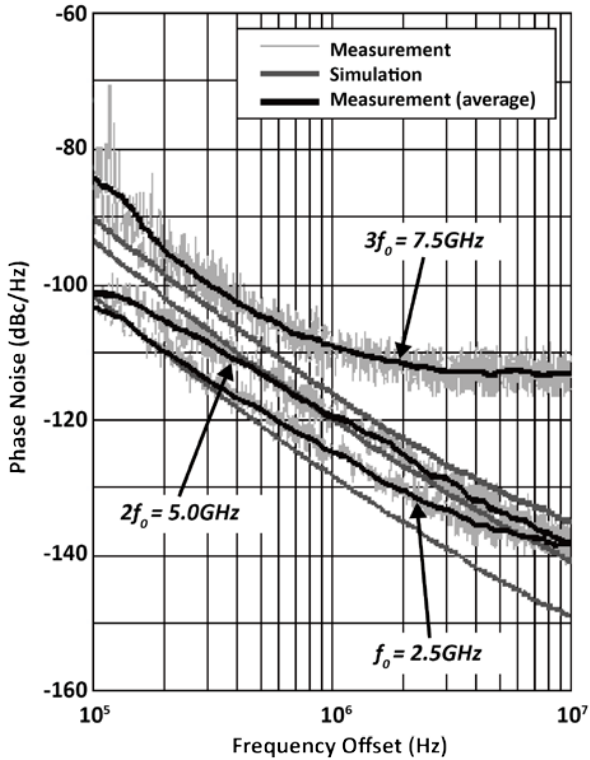

Fig. 12. Measured and simulated phase noise at 1MHz offset for 2.5, 5.0, and 7.5GHz bands.

The phase noise is measured as -125.2dBc/Hz, -119.8dBc/Hz and -109.3dBc/Hz at 1 MHz offset in the three bands, respectively. Fig. 12 shows the measured and simulated phase noise at 2.5GHz band, 5GHz band, and 7.5GHz band and at center frequency of each band's tuning curve (i.e., with a 1.2 V voltage difference across the varactors). For the first two bands, measurements show good agreements with the simulation results. For 7.5GHz band, the measured result deviates slightly from the simulation results, mainly due to the increased parasitics at higher frequency and the center frequency drift of the built-in band pass filters.

According to the measured results, a tuning range from 1.4GHz to 8.7GHz is achieved with three frequency bands. Figure 13 shows the tuning curve of 1$^{st}$ frequency band. Centered at 2.5GHz, the 1$^{st}$ band covered a tunable range from 1.4GHz to 2.9GHz with two digitally controlled bits.

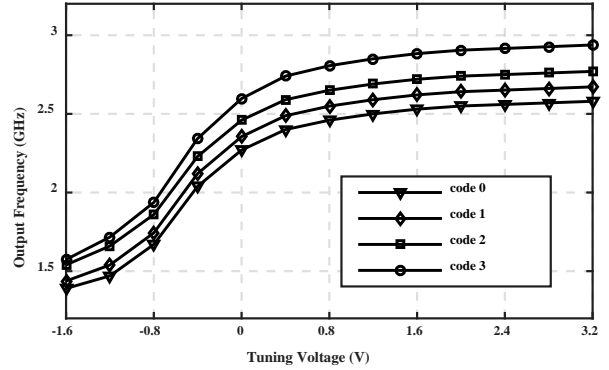

Fig. 13. 1$^{st}$ band frequency tuning curves.

Table I summarizes the performance of the implemented QVCO and the comparison with previously published VCO designs. The proposed QVCO based frequency generator achieves a FoM of 183.0dB, 183.6dB and 170.9dB for each band with the FoM defined as:

$$FoM = 10\log\left[\left(\frac{f_0}{\Delta f}\right)^2 \frac{1mW}{P}\right] - \mathcal{L}(\Delta f) \quad (15)$$

$$FoM_T = FoM - 20\log\left(\frac{FTR}{10}\right) \quad (16)$$

where $L(\Delta f)$ is the phase noise at the $\Delta f$ offset from the central oscillation frequency $f_0$, and $P$ is the QVCO's core power consumption in the unit of mW. Note that this FoM doesn't include the impact of tuning range on phase noise performance of the double frequency band was obtained without additional power, while the triple frequency band requires additional power for the mixer cell.

## V. CONCLUSIONS

An analytical model on QVCO phase noise performance is derived in this paper. It demonstrates that applying phase shift on the coupling paths can reduce the QVCO phase noise up to 9dB comparing to the same QVCO without phase shift. Implemented in a 0.18um SiGe BiCMOS technology, this paper presents a triple-band frequency generation scheme based on a hybrid BS-QVCO using BJT transistors for oscillation and NMOS devices for coupling. The triple band frequency generation RFIC covers a frequency range from 1.4GHz to 8.7GHz with three frequency bands centered at 2.5GHz, 5GHz, and 7.5GHz. The measured phase noise in each band at 1MHz offset is -125.2, -119.8 and -109.3dBc/Hz, respectively. This work demonstrated a wide tuning QVCO with a FoM$_T$ of 199.6dB.

TABLE I PERFORMANCE SUMMARY AND COMPARISON OF VARIOUS QVCOS.

| Ref. | Tech | Coupling Schemes | Power (mW) | Freq. (GHz) | Phase Noise @1MHz offset (dBc/Hz) | Tuning Range(%) | FoM (dB) | FoM$_T$ (dB) |
|---|---|---|---|---|---|---|---|---|
| [10] | 0.25 μm CMOS | Super Harmonic Coupled QVCO | 22 | 4.9 | -125 | 12 | 185.0 | 185.8 |
| [12] | 65nm CMOS | Transformer based QVCO | 6 | 4 | -121.8 | 78 | 188.2 | 203.9 |
| [13] | 0.13 μm CMOS | TS-QVCO | 10.8 | 4.5 | -120.1 | 17.8 | 184.5 | 187.0 |
| [23] | 0.35 μm CMOS | BS-QVCO | 20.8 | 2 | -140@3MHz | 17 | 183.3 | 185.6 |
| [24] | 0.13 μm CMOS | CC-QVCO | 4.2 | 5.3 | -122 | 4 | 191.4 | 187.4 |
| [25] | 0.18 μm CMOS | PTIC-QVCO* | 9 | 9.6 | -121@3MHz | 6.6 | 182.6 | 180.8 |
| [26] | 65nm CMOS | P-QVCO | 7.4 | 6.3 | -116 | 52 | 185.4 | 199.7 |
| [27] | 65nm CMOS | QVCO | 10 | 3.6 | -124.9 | 57 | 186 | 201.1 |
| [28] | 90nm CMOS | QVCO | 31 | 1.75 | -120.0 | 145 | 170 | 193.2 |
| [29] | 0.18 μm CMOS | SVCO | 9.2 | 6.3 | -124.2 | 66 | 190.5 | 206.8 |
| **This Work** | 0.18 BiCMOS | BS-QVCO | 12(43*) | 2.5[+] <br> 5.0[+] <br> 7.5[+] | -125.8[+] <br> -119.9[+] <br> -109.3[+] | 145 | 183.0[+] <br> 183.1[+] <br> 170.5[+] | 193.7~ 206.3 |

* QVCO core consumes 12mW power, while the mixer for 3rd frequency band generation consumes extra 31mW power.

[+] measured at center of each three frequency bands.